\documentclass[twocolumn,iicol,pdflatex]{sn-jnl}
\usepackage{csquotes}
\usepackage{graphicx}%
\usepackage{multirow}
\usepackage{multicol}
\usepackage{amsmath,amssymb,amsfonts}%
\usepackage{amsthm}%
\usepackage{mathrsfs}%
\usepackage[title]{appendix}%
\usepackage{xcolor}%
\usepackage{textcomp}%
\usepackage{manyfoot}%
\usepackage{booktabs}%
\usepackage{algorithm}%
\usepackage{algorithmicx}%
\usepackage{algpseudocode}%
\usepackage{listings}%
\usepackage{hyperref}
\usepackage[
backend=biber,
style=apa, 
sorting=nyt
]{biblatex}
\title[Phase ordering in the near-critical brain]{Phase ordering in the near-critical regime of the Alzheimer's and normal brain}

\author[1]{\fnm{Anirudh} \sur{Palutla}}
\equalcont{\small{Equal contribution}}

\author[1]{\fnm{Shivansh} \sur{Seth}}%
\equalcont{\small{Equal contribution}}

\author*[1]{\sur{S.S.} \fnm{Ashwin}}%
\email{ss.ashwin@gmail.com}

\author*[1]{\fnm{Marimuthu} \sur{Krishnan}}%
\email{m.krishnan@iiit.ac.in}

\affil[1]{\orgdiv{Center for Computational Natural Sciences and Bioinformatics}, \orgname{International Institute of Information Technology}, \orgaddress{\city{Hyderabad}}}

 \abstract{
 Criticality, observed during second-order phase transitions, is an emergent phenomenon. The brain operates near criticality, where complex systems exhibit high correlations. The critical brain hypothesis suggests that the brain becomes an efficient learning system in this state but poor in memory, while sub-criticality enhances memory but inhibits learning. As a system approaches criticality, it develops "domain"-like regions with competing phases and increased spatiotemporal correlations that diverge. The dynamics of these domains depend on the system's proximity to criticality. This study investigates the phase ordering properties of a spin-lattice model derived from Alzheimer's and cognitively normal subjects, expecting significant differences in their proximity to criticality. However, our findings show no conclusive distinction in the distal properties from criticality, as reflected in the phase ordering behavior of the Alzheimer's and cognitively normal brain.}


\begin{document}
\maketitle
\section{Introduction}\label{sec1}

The brain is a highly complex system with a very large number of interdependent parts which exhibit non-linearity and emergent collective behavior very similar to statistical physics models of phase transitions~\parencite{chialvo2010emergent}. Functional magnetic resonance imaging (fMRI)~\parencite{glover2011overview} is a  primary tool used to study brain activity by signals sensitive to blood flow and oxygenation at a local area of the brain. fMRI uses the ``hemodynamic response" to measure regional activity in the brain~\parencite{iadecola2017neurovascular}.
The signal detected by fMRI is not a direct measure of neural activity; however, it is an indirect measure of the hemodynamic response to neural activity. By measuring changes in blood flow and oxygenation, fMRI can provide a spatial and temporal map of brain activity, allowing one to identify which regions of the brain are active during different cognitive or perceptual tasks.

The Ising model~\parencite{baxter2016exactly} is a physical model commonly used to understand the brain. The model represents a lattice with spins located at each point, which is described by the Hamiltonian $H=\sum_{<ij>}J_{ij}S_iS_j$, where $S_i$ and $S_j$ represent the spins at lattice points $i$ and $j$ and $J_{ij}$ is the pairwise coupling between these points. The spins can take values of "+1" or "-1", and the summation is limited to its nearest neighbours (as denoted by $<>$).
This Hamiltonian captures the energetic aspects, while the entropy arises from the collective degrees of freedom of the spins.

At low temperatures, the energetics of the system dominate over entropy, resulting in spin alignment and the phase known as the ferromagnetic phase. Conversely, at high temperatures, entropy dominates, and the spins exhibit random behavior, which is referred to as the paramagnetic phase. The Ising model demonstrates predictable collective behavior and undergoes a second-order transition at a critical temperature $T_c$ between these two phases. As one approaches $T_c$, competing domains emerge due to the interplay between thermal fluctuations and spin interactions~\parencite{chaikin1995principles}. Domains are regions within the material where the magnetization is uniform and distinct from its neighboring domains.
The resting-state brain has been shown~\parencite{fox2005human,eguiluz2005scale} to display correlated and anticorrelated subnetworks,  which are dynamic and spatially distributed, precisely the signature of domains in spin models.

As the temperature approaches $T_c$, the Ising spins exhibit critical behavior characterized by power-law scaling, for example,  the magnetic susceptibility diverges as $\chi \sim |T - T_c|^{-\gamma}$, where $\gamma$ is the critical exponent for the magnetic susceptibility. Similarly, the correlation length diverges as $\xi \sim |T - T_c|^{-\nu}$, where $\nu$ is the critical exponent for the correlation length. This criticality is identified as the "self-organized" criticality of the Ising model and is known to mimic the metastable states~\parencite{chialvo2010emergent,tognoli2014metastable} of the resting brain~\parencite{das2014highlighting}. At $T_c$, the system displays scale-invariant behavior, characterized by fluctuations existing across all length scales. This behavior is described by a general scale-invariant mechanism. 

The criticality of the Ising model is of interest because it serves as a simplified model for comprehending complex systems that exhibit similar behavior, such as the brain~\parencite{chialvo2010emergent}. In the brain, neurons interact with one another, and the nature of their interaction depends on their activity levels. Like the Ising model, the brain can display critical behavior~\parencite{beggs2003neuronal,beggs2004neuronal,linkenkaer2001long}, featuring scale-invariant activity patterns~\parencite{novikov1997scale}.

To liken the voxels to lattice points in an Ising model, the fMRI time series are binarized.
The strength of spin interactions can be calculated using a maximum likelihood approach based on binary activity patterns derived from fMRI data. 
The Ising model assumes fixed coupling strength and restricts interactions to nearest neighbors. 
However, brain activity patterns are not limited to nearest neighbors, and the interactions are not uniform. A variant of the Ising model, the Sherrington-Kirkpatrick model (SK)~\parencite{sherrington1975solvable}, allows for interactions beyond nearest neighbors and incorporates a distribution of coupling strengths, similar to the interaction of fMRI signals between non-local voxels. The SK variant introduces additional physics, including a new phase known as the spin-glass phase. In this phase, the spins experience frustration and exhibit glassy behavior~\parencite{mezard1987spin}. 
By mapping fMRI signals to the SK model, one can predict collective behavior in the brain, such as functional networks and critical dynamics in the presence of disorder~\parencite{ezaki2019critical}. However, it remains unclear whether any glassy features observed in the fMRI signals resemble those exhibited by the SK model.

Chialvo and Dante~\parencite{chialvo2010emergent} conjecture that, criticality is a crucial aspect of the learning and memory capacity of the brain. A brain that is sub-critical can be seen as a simple equilibrium state that is too simplistic to learn and respond effectively, while a brain that is critical has long-range correlations and small fluctuations that can bring about global changes in the neuronal patterns, which makes it a good learning system but poor in memory capacity. It is likely that the brain exists or tunes itself between these two regimes to achieve optimal efficiency. Clinical relevance to brain criticality has been an area of intense research~\parencite{zimmern2020brain}.

Alzheimer's disease (AD) is a neurodegenerative disorder that affects memory and cognition. Since cognition relies on the production and synchronization of neuronal signals~\parencite{breakspear2002detection}, studying the collective behavior of these signals is an appropriate way to investigate AD. Previous studies have explored whether AD exhibits deviations from criticality. In normal individuals, synchronization in electroencephalography (EEG) shows power-law scaling~\parencite{stam2004scale}. In individuals with AD, EEG also exhibits power-law behavior, but with decreased amounts in certain frequency, regimes compared to non-demented patients~\parencite{stam2005disturbed}. The power-law exponents of the spectral densities showed statistically significant differences between AD and control subjects in the temporal and frontal lobes~\parencite{vyvsata2014change}, which is consistent with frontal lobe atrophy associated with AD. Magnetoencephalography (MEG) studies, which infer magnetic fields produced by brain electric currents, showed decreased autocorrelations and oscillation bursts in the signals compared to controls~\parencite{montez2009altered}.

 How does the near-critical phase behavior or phase ordering of the fMRI in  Alzheimer's disease compare to the normal brain? The main purpose of our manuscript is to investigate quantitative spatiotemporal features of the so-called domains near the criticality of the normal and Alzheimer's brain. We study the fMRI signals for both normal human brains and Alzheimer's brains in a resting state, and we characterize features from the domain properties that are similar and distinct in these cases.

 \section{Materials and methods}
 
\begin{figure*}[h]%
\centering
\includegraphics[width=0.8\textwidth]{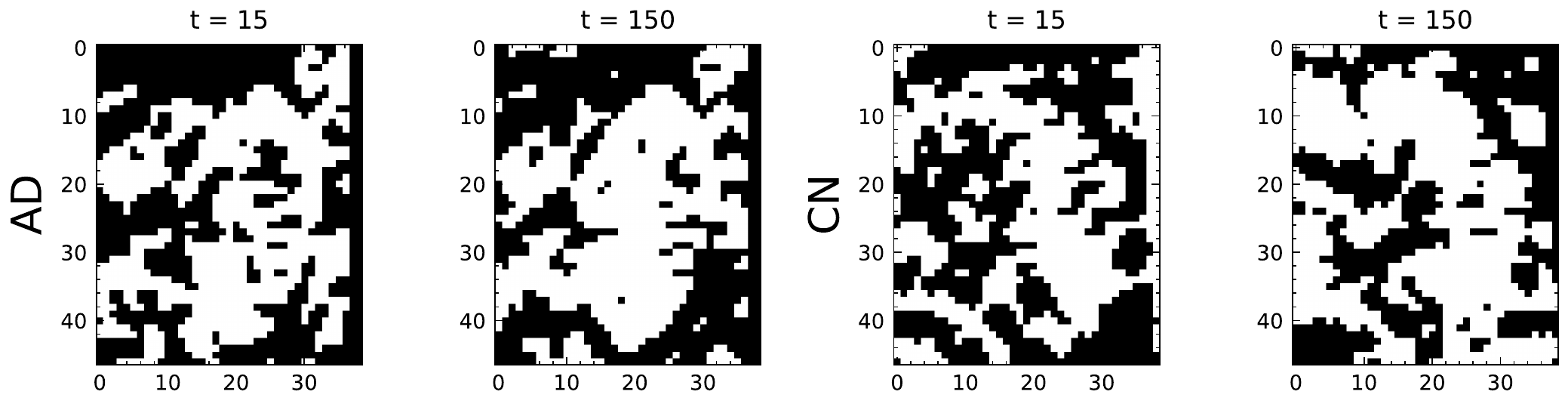}
        \caption{Domain formation in near critical brain of Alzheimer's (AD) and control (CN) subjects for $t$=15 secs and $t$=150 secs.}\label{domain}
\end{figure*}
 \subsection{Data acquisition}
 The study utilized resting-state functional magnetic resonance imaging (rs-fMRI) data from the Alzheimer’s Disease Neuroimaging Initiative (ADNI) database \parencite{petersen_alzheimers_2010}. To obtain rs-fMRI data for Alzheimer’s patients (AD) and cognitively normal (CN) subjects, we used the portal:~\url{adni.loni.usc.edu}. We found 121 AD images and 243 CN images on the portal. However, since ADNI contains multiple rs-fMRI scans for some subjects at different time points, we only selected one scan per subject. Our final data-set consisted of rs-fMRI scans for 89 subjects, including 34 AD and 55 CN subjects. The corresponding anatomical scans were also used during preprocessing.
 
 \subsection{Preprocessing}
The fMRI data was preprocessed primarily using tools from the FMRIB Software Library (FSL) \parencite{fsl-main-1, fsl-main-2}. First, motion correction was performed using FSL's MCFLIRT \parencite{mcflirt} to align all volumes to the mean volume, producing motion parameters and mean images as output. Next, FSL's SliceTimer was used for slice-timing correction. The coregistration step included the following procedures: (1) skull-stripping the anatomical image using FSL's BET, (2) segmenting it with FSL's FAST and thresholding the resulting white matter probability image, (3) pre-alignment and coregistration of the fMRI to the anatomical images using FSL's FLIRT, and (4) applying the computed coregistration transformation to the functional and mean images. 
The images were spatially smoothed using SPM with a full-width at half-maximum (FWHM) of 5mm. 
Nipype's ArtifactDetect algorithm \parencite{nipype-alg-ref} was used to detect and separate out artifacts from the functional images, with a norm threshold of $2$ and z-intensity threshold of $3$. 
Finally, Nilearn~\parencite{nilearn-alg-ref} was used to calculate and apply a brain mask, which utilized the histogram of the mean fMRI image intensity and discarded the bottom $20$\% and top $15$\% of it.

\subsection{Correlations}
\subsubsection{Self Averaging}


To investigate self-averaging, we calculate $R_X=(\Delta X)^2/[X]^2$, which depends on system size $N$.
Here, $X$ is a random variable taken from a distribution $P(X)$. The fMRI data for a subject is taken as a flattened array of voxel strengths at each time point. Systems of different sizes are created from this fMRI data array by randomly selecting non-overlapping subarrays of size $N$, and the mean of each subarray are calculated. The $R_X$ value of each new system is calculated and plotted against $N$ on a log-log scale. This process is repeated for systems of sizes 1 to 1000 and for all subjects at a fixed time.

\subsubsection{Time correlation}
The auto-correlation function (ACF), which is denoted as $\rho(t)$, of the BOLD signal is calculated using the inverse Fourier transform power spectral density (PSD) and ACF as defined by Wiener-Khinchin theorem~\parencite{chatfield1989analysis}. 
$\rho(t)$ is normalized by the value at $t=0$. A stretched exponential function of the form given below was fit to the ACF using the least squares method.

\begin{align}
\rho(t)= A\exp\left[-\left(\frac{t}{\tau} \right)^\beta\right] + B
\end{align}
$A$ is chosen to be unity and $B$ is chosen to be the average value of the second half of the time series. Finally, the relaxation time $\tau$ and the stretching parameter $\beta$  are extracted from the fit. 


 \section{Results and Discussion}
Our investigation focuses on the domains present in fMRI signals, which we analyze by associating them with a spin system. To do this, we first assign 3D lattice points to the centers of the voxels. Then, we map the fMRI signals in each voxel to spins. This involves finding the average of the highest and lowest ten signals and designating those values as the maximum and minimum signals for that voxel. We then create a linear map which assigns +1 and -1 to the maximum and minimum signals respectively. Positive values in this new signal are mapped to ``+1" and negative values to ``-1". Note, the threshold for each signal is sometimes taken to be the mean of the signal, however, this restricts the spin model to the paramagnetic phase. The above method for calculating the threshold avoids this issue.

By connecting neighboring spins on the lattice that have the same spin with an edge, we define a 'domain' as the resulting set of lattice points connected by these edges. Domains are represented by black and white regions in Fig. (\ref{domain}), where white indicates +1 spins and black indicates -1 spins. 
Our analysis reveals the presence of large percolating domains of magnetization with +1 and -1 spins in both the AD and CN cases. 
The existence of such large domains suggests that the system is close to the critical temperature ($T_c$) for the subjects studied~\parencite{chaikin1995principles}.
To compare AD and CN cases quantitatively, we identify domains for all the time series and subjects in our study. We study the distributions of the number and size of domains, represented as $P(n_{dom})$ and $P(S_{dom})$, respectively. $S_{dom}$ is the number of spins contained in a domain and $n_{dom}$ is the number of domains. To identify domains assigned to the voxels across the signal time series for each subject, we use the Hoshen-Kopelman algorithm~\parencite{hoshen1976percolation}. We compute $P(n_{dom})$ and $P(S_{dom})$ throughout the time series of length $T_{max}$, considering all voxels $N_{vox}$ and all subjects $N_{sub}$:
\begin{align}
    P(S_{dom}) = \frac{1}{N_{sub}T_{max}}\sum_{j}^{N_{sub}}\sum_{k}^{T_{max}} \delta(S_{dom} - S^{j}(k))
\end{align}
\begin{align}
    P(n_{dom}) = \frac{1}{T_{max}}\sum_{k}^{T_{max}} \delta\left(n_{dom} - \frac{1}{N_{sub}} \sum_{j}^{N_{sub}} n^{j}(k)\right)
\end{align}

\begin{figure*}[h]%
\centering
\includegraphics[width=0.8\textwidth]{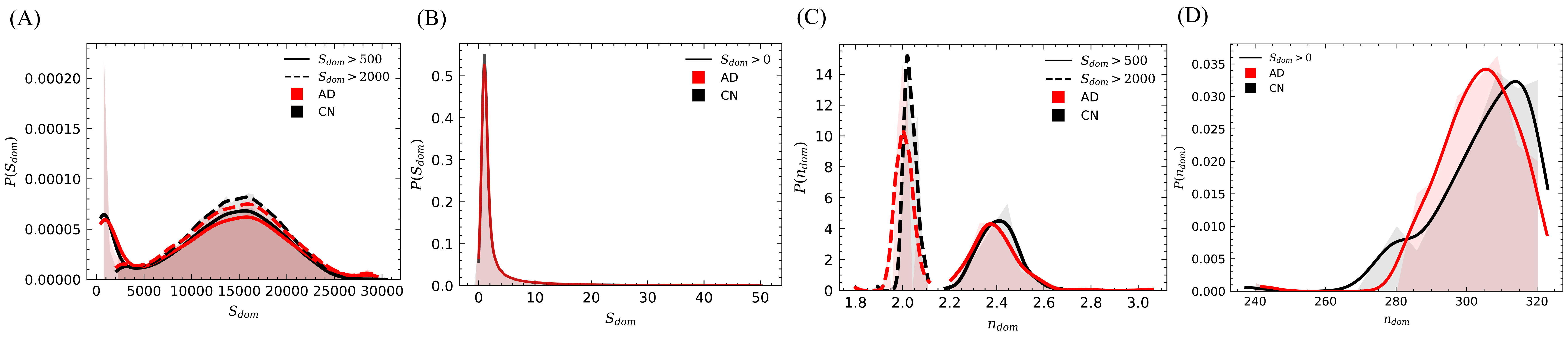}
        \caption{
        (A,B) Domain size distributions $P(n_{dom})$ for all subjects throughout the time-series limited to domains of size (A) $S_{dom}>500,2000$ and (B) $0<S_{dom}<50$. (C,D) Distribution of number of domains $P(n_{dom})$ averaged over all subjects (at each time point) limited to domains of size (C) $S_{dom}>500,2000$ in and (D) $S_{dom}>0$.
        }
        \label{PnPs}
\end{figure*}

Here, $n^{j}(k)$ and $S^{j}(k)$ represent the number of domains and size at time $k$ for subject $j$.
Fig. \ref{PnPs} (A) shows the $S_{dom}$ distribution for large domains ($S_{dom} > 500, 2000$). In Fig. \ref{PnPs} (B) we find that smaller domains ($S_{dom} < 50$) are dominated by sizes $\lessapprox$ 10 domains. 
Since the distribution is very sparse, relevant domain size limits were set to study the large-domain region (in (A) and (C)) and the small-domain region (in (B) and (D)).
There appear to be no significant differences in cluster sizes between AD and CN for both small and large domains.
Our investigation shows large, percolating domains the time series for every subject.  This can be seen in theh peak in (A) at $S_{dom} \approx 15000$, corresponding to the peaks at $n_{dom}\approx 2.0$ and $n_{dom}\approx 2.4$ in (C).
Interestingly, there also seem to be small, isolated domains which can be seen in the initial peak in (B), corresponding to the peak at $n_{dom} \approx 300$ in (D). 
This seems to imply the existence of a couple of stable clusters with the dynamics mostly revolving around smaller domains.
 
Self-averaging is a fundamental concept in thermodynamics, indicating that the statistics of a system improve with an increase in system size. According to the central limit theorem (CLT), fluctuations become proportional to $N^{-1/2}$, where $N$ is the system size. However, the CLT assumes independence in the random variables whose average is being calculated. When individual components of a system evolve differently due to increasing correlation lengths, self-averaging tends to break down. As a system approaches criticality, the domain sizes tend to increase because individual domains have different phases, leading to a breakdown of self-averaging.
Self-averaging, or the lack thereof, can be quantified \parencite{mezard1987spin,aharony1996absence,roland1989lack}. Let's consider a random system where an observable property takes on random variables from a distribution, $P(X)$ with variance, $\Delta V=[X^2]-[X]^2$, and its average, $[X]$ (averaged over realizations of the randomness). We can define a quantity, relative variance $R_X=\Delta V/[X]^2$. According to the CLT, when $R_X\sim N^{-1}$, we say the system is self-averaging. However, when $R_X\sim N^{-\alpha}$ and $0<\alpha<1$, self-averaging is poor.

\begin{figure*}[h]%
\centering
\includegraphics[width=0.8\textwidth]{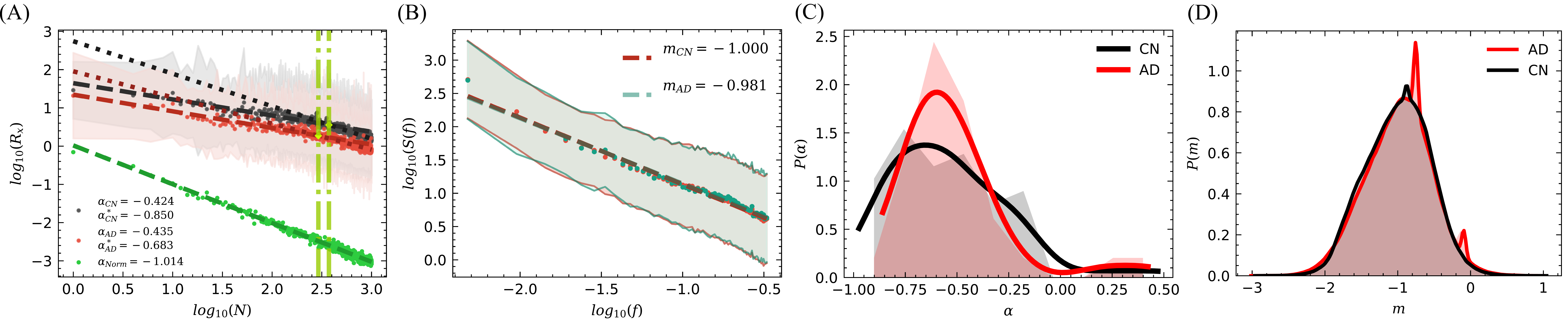}
        \caption{    
        (A) Log-log plot of $R_X$ vs $N$ is shown for mean over the control (CN), Alzheimer (AD) subjects and random variables taken from a normal distribution (Norm). The scatter points are mean values over subjects and the shaded regions show the standard deviation. The dashed lines represent a fit on the first $50\%$ of data points, and the dotted lines are fit on the final $50\%$ of the data. Their intersection point (fluorescent) $N^{\ast}_{CN} \approx 372$, $N^{\ast}_{AD} \approx 291$ marks $N$ after which system moves closer to self-averaging. 
        (B) Log-log plot of $S(f)$ for CN and AD. Dashed lines show the mean linear fit and shaded regions show the standard deviation of $S(f)$. 
        (C) As $R_X \sim N^{\alpha}$, the distribution of $\alpha$: $P(\alpha)$ derived from (A) is shown for the CN and AD case. 
        (D) The distribution $m$, the slope of $S(f)$
        }
        
        \label{tau}
\end{figure*}

To examine self-averaging, we calculate $R_X$ for fMRI data in the CN and AD cases. Fig. \ref{tau}(A) shows the distribution of these alpha values. 
Notably, in the log-log plot, $R_X$ exhibits two distinct slopes, transitioning from $\alpha=-0.424$ to $\alpha=-0.850$ at $N^*\approx 372$ for CN and from $\alpha=-0.435$ to $\alpha=-0.682$ at $N^*\approx291$ for AD. $N^*$ is the intersection point between the linear fits for the first $50\%$ and last $50\%$ of the data points. The magnitude of the slope of the last $50\%$ for AD ($0.682$) is significantly lower than that of CN ($0.850$) implying worse self-averaging in the AD case.
For comparison, we also plot $R_X$ for the normal distribution with an $\alpha\approx-1$. The poor self-averaging observed indicates criticality in the presence of disorder. $N^*$ depends on how close the system is to criticality. The significance of $N^*$ becomes apparent when statistics involve voxel averaging. $N^*$ depends on how close the system is to criticality.  In fMRI studies, due to the substantial spatial resolution of the signals, it is common to reduce voxel-wise data to a few hundred regions of interest (ROIs) based on pre-existing atlases~\parencite{roland_brain_1994}.
This is done through a process called parcellation where each voxel is mapped to an existing anatomical or functional parcel/ROI. The time-series is obtained by averaging over the voxels present within a parcel. Our analyses indicate that parcellation with ROIs smaller than $N^*$ would lead to inadequate self-averaging. From the case studied here, the lower limit for the number of voxels in a ROI seems to be around 400.


We calculate the power spectrum $S(f)$ for brain signals versus frequency ($f$) in the AD and CN cases, as shown in Fig. \ref{tau} (B). The power spectrum $S(f)$ follows a power-law relationship $S(f) \propto f^m$, where the exponent $m$ characterizes the color of the noise. A power spectrum that follows $S(f) \propto 1/f$ indicates self-similarity and modular hierarchical organization in the brain~\parencite{expert2011self}. We observe the $1/f$ behavior in both AD and CN cases, with mean exponents ranging from $-0.98 \pm 0.45$ for AD to $-1.00 \pm 0.44$ for CN. The distribution of $m$ is depicted in Fig. \ref{tau} (D) and is similar for both AD and CN, suggesting that the hierarchical self-similar organization may not differ significantly between the two cases.
        
\begin{figure*}[h]%
\centering
\includegraphics[width=0.8\textwidth]{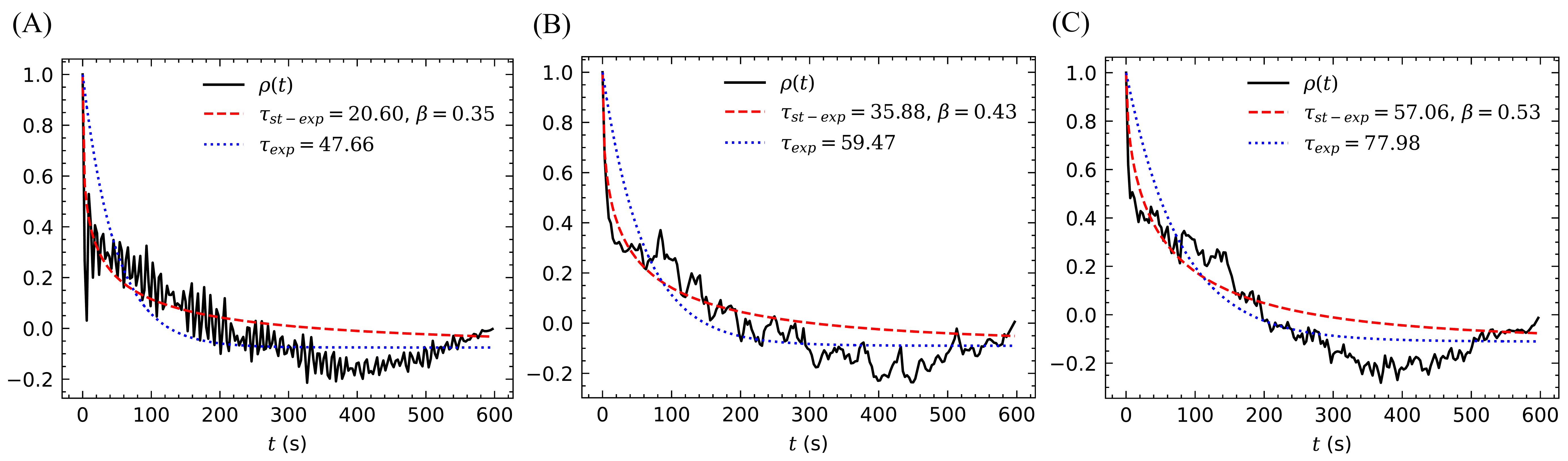}
        \caption{(A),(B),(C) Time correlation $\rho(t)$. The time correlations fit stretched exponential functions $\rho(t)=\exp[-(t/\tau)^\beta]$ better than pure exponentials.  }\label{fig2}
\end{figure*}
\begin{figure*}[h]%
\centering
\includegraphics[width=0.8\textwidth]{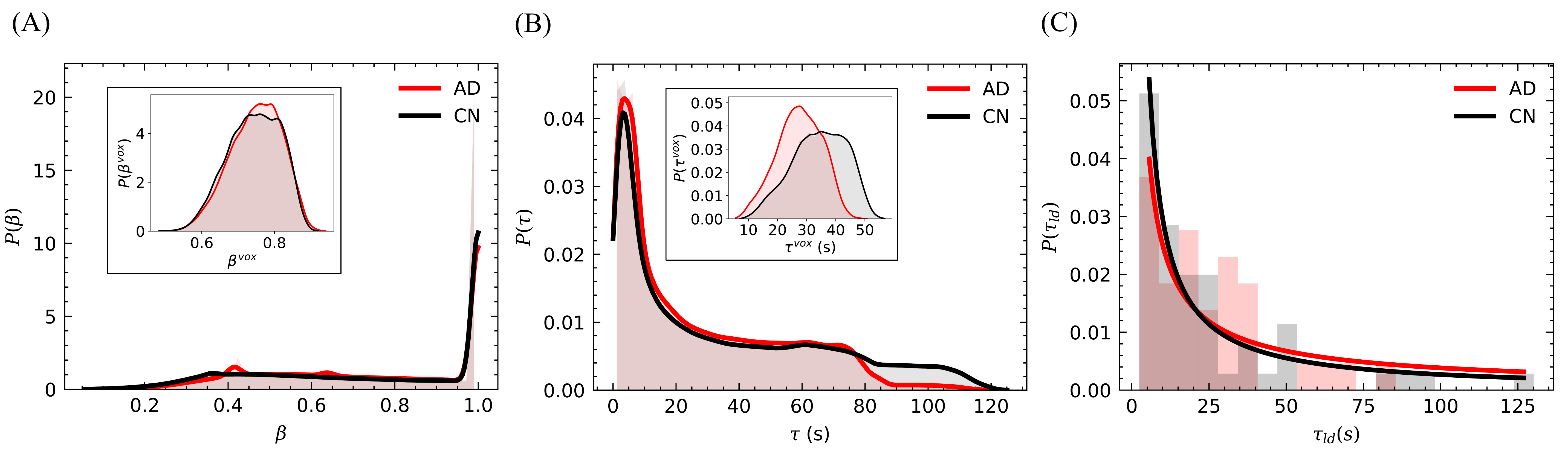}
        \caption{
        (A) Distribution of all $\tau$ values and (inset) $\tau$ values averaged over subjects for each voxel. (B) Distribution of all $\beta$ values for the stretched exponential fit and (inset) $\beta$ values averaged over subjects for each voxel. 
        (C) Distribution of the mean time of largest domains, $\tau_{ld}$ (in seconds) averaged over all subjects.
         }
        \label{fig5}
\end{figure*}
We investigate the time correlation function of the fMRI time series generated from each voxel. The power spectrum is calculated for each time series using the Wiener-Kinchin theorem \parencite{chatfield1989analysis}, which allows for the direct calculation of the auto-correlation through a straightforward Fourier transform of the power spectra (see Methodology for details).

Figure \ref{fig2}(A-C) shows the extracted time correlation function, $\rho(t)$. Notably, $\rho(t)$ exhibits a better fit to a stretched exponential function. Stretched exponential relaxation, also known as the Kohlsrauch-Williams-Watts stretched exponential form \parencite{KWW}, involves a two-step relaxation pattern observed in the SK model \parencite{billoire2011dynamics} and is a classic signature of metastable states approaching the glass transition.

To accurately calculate the relaxation time, $\tau$, we fit $\rho(t)$ to both an exponential decay function  and a stretched exponential decay function ($\rho(t)$) using the least squares method. Figure \ref{fig5}(A) displays the distribution of $\beta$ and $\tau$ values calculated in the following manner:
\begin{align}    
P(\beta) &= \frac{1}{N_{sub}N_{vox}}\sum_{i}^{N_{sub}}\sum_{j}^{N_{vox}}\delta\left(\beta-\beta^{i}_{j}\right)
\\   
P(\tau) &= \frac{1}{N_{sub}N_{vox}}\sum_{i}^{N_{sub}}\sum_{j}^{N_{vox}}\delta\left(\tau-\tau^{i}_{j}\right)
\end{align}
Here, $\beta^i_j$ and $\tau^i_j$ are the values of $\beta$ and $\tau$ for $j^{th}$ voxel and $i^{th}$ subject. 
We observe that approximately 59\% of voxels have $\beta < 0.95$, indicating a deviation from exponential decay. AD and CN show no significant differences in $P(\beta)$. However, $P(\tau)$ shows (Fig. \ref{fig5}(B)) difference between AD and CN only in the large $\tau$ range ($\tau > 80$ ). The inset in Figure \ref{fig5}(A, B) shows a different way of averaging, specifically averaging over the voxels
\begin{align}
P(\beta^{vox}) &= \frac{1}{N_{sub}}\sum_{i}^{N_{sub}}\delta\left(\beta-\frac{1}{N_{vox}}\sum_{j}^{N_{vox}}\beta^{i}_{j}\right)
\\
P(\tau^{vox}) &= \frac{1}{N_{sub}}\sum_{i}^{N_{sub}}\delta\left(\tau-\frac{1}{N_{vox}}\sum_j^{N_{vox}}\tau^{i}_{j}\right)
\end{align}

$P(\beta^{vox})$ shows that the entire distribution follows a stretched exponential pattern, and there are no significant differences between AD and CN cases. On the other hand, $P(\tau^{vox})$ exhibits a clear distinction between the AD and CN cases, indicating that CN is closer to the critical temperature than AD.
The presence of a stretched exponential relaxation suggests that some parts of the lattice may be in a spin-glass phase, contributing to the increased complexity of criticality compared to the Ising model \parencite{AT}. 
Furthermore, we track the largest domain $S_{ld}(t)$ over time by finding the domain which has maximal overlap with the largest domain at the previous time point, this is done over all time points. 
We then calculate the autocorrelation function $<S_{ld}(t)S_{ld}(0)>$ for each subject. The autocorrelation shows an exponential decay. We obtain relaxation times $\tau_{ld}$ using an exponential fit. Figure \ref{fig5}(C) displays the distribution of relaxation times for the AD and CN subjects. We observe that the relaxations also exhibit an exponential decay, and the distribution of $\tau_{ld}$ is similar for both AD and CN.

\section{Conclusion}
We have conducted an investigation into the phase ordering domains and critical dynamics of Alzheimer's disease (AD) and cognitively normal (CN) individuals. Our analysis revealed that both cases exhibit characteristics of being near critical, but we were unable to definitively determine which case is closer to criticality. However, an examination of relaxation times suggests that CN may be slightly closer to criticality, as indicated by a shift in the tail towards larger values in the distribution of mean relaxation times. At the critical temperature ($T_c$), the relaxation time tends to diverge, so a longer relaxation time indicates a closer proximity to $T_c$. However, this does not seem to hold true in terms of domain sizes, as the domain sizes of CN are not significantly larger than those of AD. This leads us to believe that criticality in the brain may not be as straightforward as the criticality observed in the Ising model.
We observed that both AD and CN exhibit time correlation functions with stretched exponential features, similar to those found in the spin-glass phase. This suggests that the domain ordering in the brain may possess characteristics akin to the SK model.
Traditionally, the brain has been regarded as a near-critical system capable of adjusting its criticality to optimize learning and memory \parencite{chialvo2010emergent}. However, this perspective assumes that the criticality follows the Ising class. In spin-glass phases, the presence of large barriers makes it difficult for the system to escape, resulting in persistent memory traits. 
Our findings indicate that the nature of criticality in the brain is more likely of the spin-glass type~\parencite{ezaki2019critical}, offering a broader range of complex features to explore \parencite{AT}. Nonetheless, even this does not provide an explanation for why the phase ordering dynamics observed in the cases studied do not display significant deviations. Our research prompts further investigation into the assertions regarding the brain's efficiency and its distant properties from criticality \parencite{o2022critical}.

\section*{Declarations}
\subsection*{Competing Interests}
The authors have no competing interests to declare that are relevant to the content of this article.

\printbibliography
\end{document}